# Technological Developments and Factor Substitution in a Complex and Dynamic System

Loet Leydesdorff

Science & Technology Dynamics, University of Amsterdam

Amsterdam School of Communications Research (ASCoR)

Kloveniersburgwal 48, 1012 CX  Amsterdam, The Netherlands

and

Peter Van den Besselaar

Social Science Informatics, University of Amsterdam

Roetersstraat 15, 1018 WB  Amsterdam, The Netherlands

**Abstract**

Schumpeter's (1939) distinction between changes in the form of the production function corresponding to innovation, and shifts along the production function corresponding to factor substitution, does not preclude that the underlying dynamics interact.  In an evolutionarily complex system, such interactions are expected: they lead to non-linear terms in the model, and therefore to stabilization and self-organization in addition to selection and variation.  Relatively simple simulations enable us to specify various concepts used in "evolutionary economics" in terms of non-linear dynamics.  While a technological trajectory can be considered as a stabilization in a (distributed) environment, a technological regime can be defined as a next-higher-order globalization in a hyper-space.  A regime is able to restore its order despite local disturbances, for example by the political system.  Technology policies may be effective at the level of the (sub-)systems if they provide the relevant agents with room for "creative destruction" of the globalized hyper-systems.  Implications for firm behaviour and innovation policies are elaborated.

# TECHNOLOGICAL DEVELOPMENTS AND FACTOR SUBSTITUTION
# IN A COMPLEX AND DYNAMIC SYSTEM

**Introduction**

The understanding of factor substitutions in relation to technological developments is crucial to major problems in advanced industrial economies, like (un-)employment and pollution. In evolutionary economics, technological regimes and technological trajectories have been distinguished for understanding the results of non-linear dynamics between the price mechanism and innovation. The theory of self-organizing systems renders a new perspective on the dynamic interactions among these mechanisms.

In this study, we provide operational definitions for the various concepts that have been proposed in evolutionary economics, by using simulations. The starting point of our simulations is Schumpeter's (1939, at p. 87) distinction between shifts along the production function and changes in the form of the production function indicating technological innovations. Although the later Schumpeter (1966) emphasized that technological developments are also driven by economic factors (e.g., factor prices), the analytical distinction has been fruitful for both empirical research (e.g., Solow 1957; Salter 1960) and theoretical developments in this domain (e.g., Sahal 1981). However, the innovative shift of the production function could not be explained in the neo-classical model; technological progress was considered as an external given or a residual factor.

In a seminal study, Nelson and Winter (1982) proposed to make technological developments endogenous to economic theorizing by using Markov chain models.[1] These authors, however, changed also the unit of analysis. While the (neo-classical) price mechanism and (Schumpeter's) innovative dynamics had been attributed to the economy as a system (cf. Alchian 1950), Nelson & Winter (1982) focused on explaining technological developments in terms of longitudinal developments in *firm behaviour* (Andersen 1994).

Our analysis will be pursued at the level of the economic system. We shall show that the specification of boundary conditions to the price mechanism is sufficient to explain the emergence of trajectories in an economy. While the economy initially offers opportunities to local actors, organizations (e.g., corporations) are able to develop institutional mechanisms (e.g., "heuristics") for optimizing their structures with reference to two dynamics in their environment (i.e., markets and technologies; cf. Barras 1990). Nelson & Winter's (1977 and 1982) "search and selection processes" can in this model be considered as interacting procedures at the organizational interface between technological options ("search processes") and market opportunities ("selection processes").

---

1. A system is attributed the Markov property if knowledge of previous states of the system (at $t < t_0$) does not add to the quality of the prediction of the next stage (i.e., at $t = t_1$) on the basis of the current state (i.e., at $t = t_0$). In a first-order Markov chain model the matrix of transition probabilities is assumed to be constant.

By definition, functions of the social system have to be carried by actors, but at the level of the system the distribution of actors (in this case, firms or aggregations of firms into industries) can be considered as another context that provides the system with complexity in an additional dimension (cf. Leydesdorff 1993). This third dimension can be used for selections upon first-order selections (on the variation in a first dimension) using a second-order cybernetics. In other words: which selections will tend to be selected during stabilization?

The operation of two selective mechanisms upon each other in a three-dimensional systems allows for stabilization. Accordingly, a trajectory develops in relation to three relevant contexts: (i) the economic price mechanism, (ii) the `upsetting' technological innovations, and (iii) the organizational interface, i.e., firm behaviour. Since a trajectory is localized in the third dimension of distributed firm behaviour, it will over time experience another feedback from this context. This additional feedback provides us with the conditions for the emergence of a globalized "technological regime" in a four-dimensional hyper-space.

Dosi (1982, at p. 154) once used the metaphor of a "cylinder" in a multi-dimensional space for describing a technological trajectory and its trade-offs. This metaphor still appeals to a three-dimensional representation, and the focus is correspondingly on the trajectory. The technological regime (or "technological paradigm" as Dosi (1982) prefers to call it) is uncertain in a fourth dimension with respect to the "cylinder" that it will consider as the representation of its past. In other words, the higher-order system contains a degree of freedom that enables it to make additional selections on lower-order selections. However, the hyper-cycle remains by definition "absent" if one studies observable events and relations (cf. Giddens 1979). One needs an algebraic instead of a geometrical representation for the description of developments in four dimensions (Andersen 1992).

A regime is not an observable identity, but a contingent set of options for the further development of the complex system. It is based on a specific resonance among its constitutive cybernetics. Noise is continuously filtered out by the specific resonances among the lower-order systems (Simon 1973). The self-organizing regime remains subject to the price mechanims, technological learning and organizational learning (e.g., scale effects) as its constitutive cybernetics, but it performs its own cycle, and thereby transforms the economy.

**Towards a Model**

In (first-order) Markov chain models technological change is represented as a matrix of transition probabilities. Selection environments dynamically condition this probability distribution (Nelson & Winter 1977, at pp. 61 ff.). Thus, two dynamic principles are distinguished: the equilibrium function or the price mechanism corresponds in this representation to the conditional part of the transition probabilities, while new technologies "continually are introduced to upset the movement towards equilibrium" (*ibid.*, at p. 48; cf. Fels 1964 , at p. 428). While Nelson & Winter have specified the conditional relations between the two dynamics, they did not specify the dynamic interaction terms, since they were first concerned with building "conformable sub-theories" (*ibid.*, p. 49). As noted, the assumption is that corporations organize the interaction.

Economic historians, however, have emphasized the *interactive nature* of the relations among technologies and markets at the system's level (e.g., Rosenberg 1976). Sociologists have stressed the interdependence of variation and selection in social development processes (e.g., Pinch and Bijker 1984; Van den Belt and Rip 1987), and others (e.g., Sahal 1981; Arthur 1988) have correspondingly emphasized the *path-dependent* nature of technological developments.

The idea that technological developments involve an "internal momentum" (e.g., Winner 1977)--since each development extends from a previous stage--and, secondly, an extent to which each development interacts with relevant environments. The variation in the relation to an environment can become "locked" into the self-referential system if the noise surpasses a selective threshold (cf. David 1985; Arthur 1988). If such a "lock-in" of (originally external) variation into the selecting system occurs, the latter internalizes this context as another degree of freedom (Leydesdorff 1994). This procedure can be reiterated. For example, we shall argue below that a trajectory can be considered as the result of a "lock-in" between the dynamics of innovation and the market environment, and that the resulting dynamic pattern can recursively enter into another relation to its environment.

The recursive and interactive terms invalidate the use of first-order Markov chain models.[2] While the first-order Markov chain tends towards a single steady state or follows a "natural trajectory," more than one trajectory can be developed if the transition matrix for the system is allowed to vary with selection environments and/or over time. For example, Allen (1988) showed by using parallel and distributed computing, that in the curve of fish ("capital") against fishing ("labour") various densities could be originated in the simulation. Note that Nelson & Winter's (1977) concept of a "natural" trajectory is thus a consequence of their decision to first disregard higher-order terms in the Markov chain model.

---

2. In (first-order) Markov chain models, the probability distribution of the transitions is considered to be independent of the state of the system. The feedback, however, introduces a dependency on the historical state of the system.

If certain trajectories offer specific advantages to enterpreneurs, one expects the production system to be increasingly "localized" in this area of the production function (e.g., Atkinson & Stiglitz 1969; Sahal 1981). The local advantages may then begin to function as a specific "selection environment" into which the further development of this technology can get "locked." We shall show below that the co-evolution of a specific combination of a selection environment and a technological trajectory can lead to *a technological regime* or--in the case of an opposite sign--to the subsequent disintegration of this configuration.

**Baseline for the simulation**

The simple Cobb-Douglas production function depicted in *Figure One* is used as the starting configuration for our simulations. Sixteen relevant technologies are conveniently defined along this function, which itself is based on an output ($Q = K * L$) of ten units.

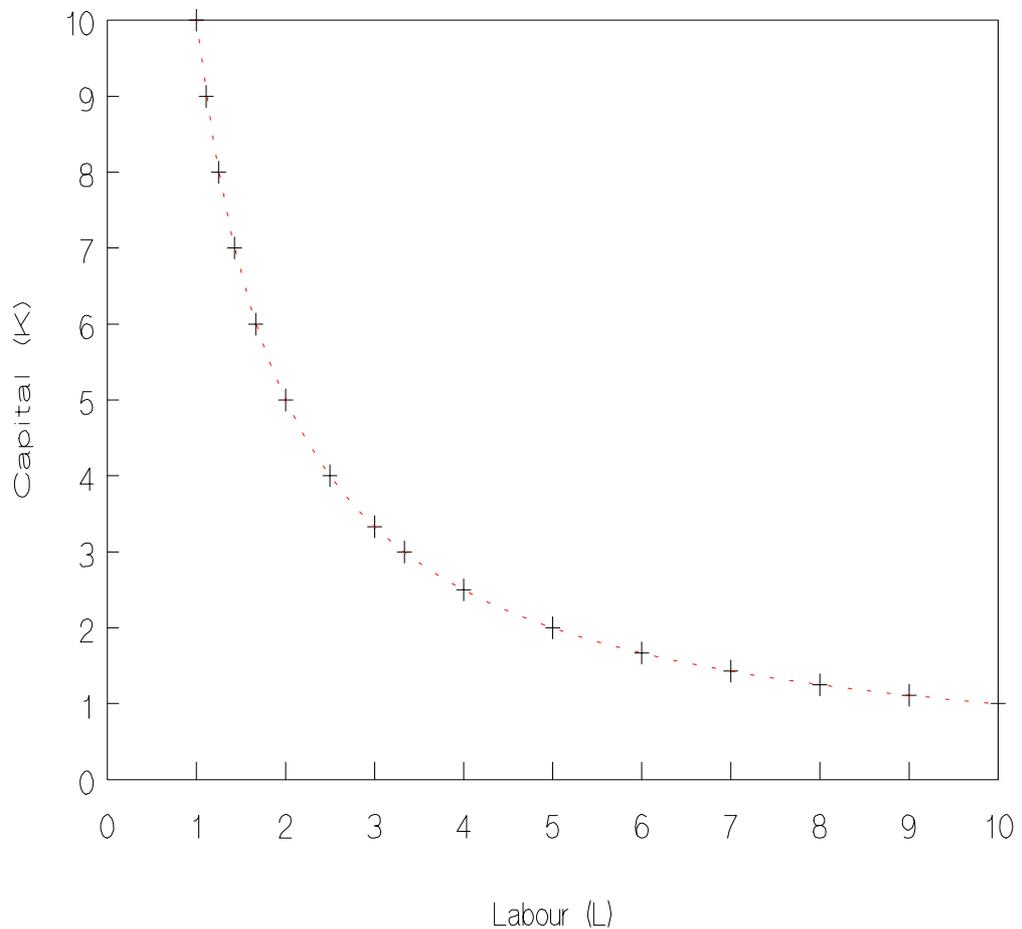

**Figure 1**

*Cobb-Douglas production function ex ante (Q = K * L = 10)*

If one simulates under the condition of factor-neutrality with a technological development of 1% reduction of factor input per period, then the two functions indicated in *Figure Two* are generated after 50 and 250 time periods respectively. This picture exhibits the well-known shift towards the origin when factor neutrality of technological developments is assumed. We shall use this figure below as a baseline for the comparison.

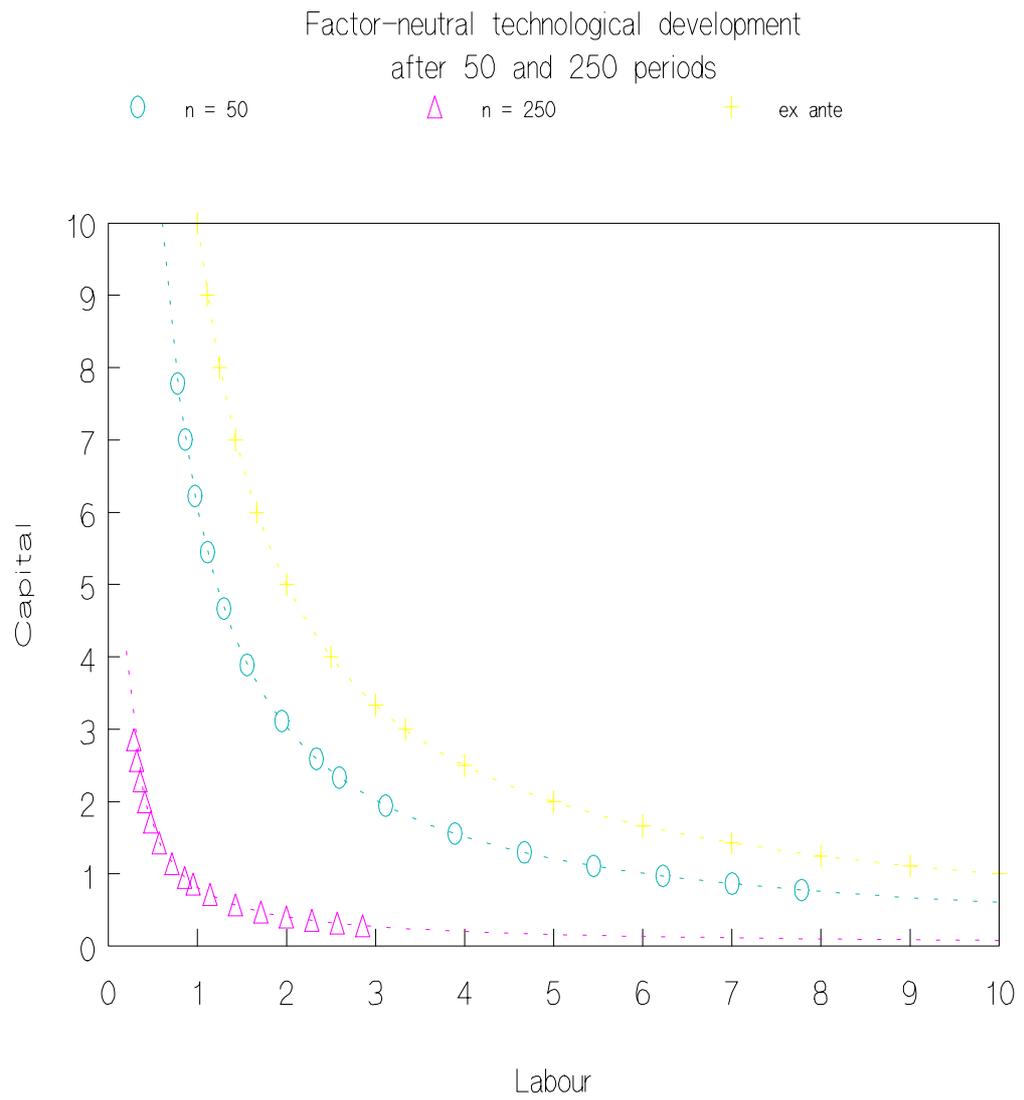

**Figure 2**

*Factor-neutral technological developments after 50 and 250 periods*

**Interaction between factor substitution and technological development**

The assumption of factor neutrality in technological developments has often been challenged (cf. Hicks 1932; Kennedy 1964). Let us assume in the next simulation that this thesis is justified for the area in which the input in terms of capital is approximately in balance with the input in terms of labour, keeping in mind that technologies which are heavily biased in terms of one of these inputs are under pressure to economize exclusively on the most intensively used input (cf. Rosenberg 1976). In this case, the production function can be divided into three regions: one with a relatively high input in capital, one with relatively high input in labour, and one of a mixed type.

One may expect the three regions to differ in terms of how the various input factors are substituted in the case of technological developments. Let us assume that, for example, in the two regions which represent *im*balanced situations, all reduction of factor costs is realized with respect to the intensively used factor, and that the imbalance reinforces technological development in the respective dimension with another percent per period. As real-life examples, one might think of a chemical plant that is highly capital-intensive, or a situation where there is plenty of labour but capital is very scarce (as in some less developed countries). Although these relatively simple assumptions may seem extreme, they enable us to show an effect that is always present when there is some form of interaction between factor substitution and technological development, notably that the smoothness of the curve disappears, and an economic rigidity may be generated.

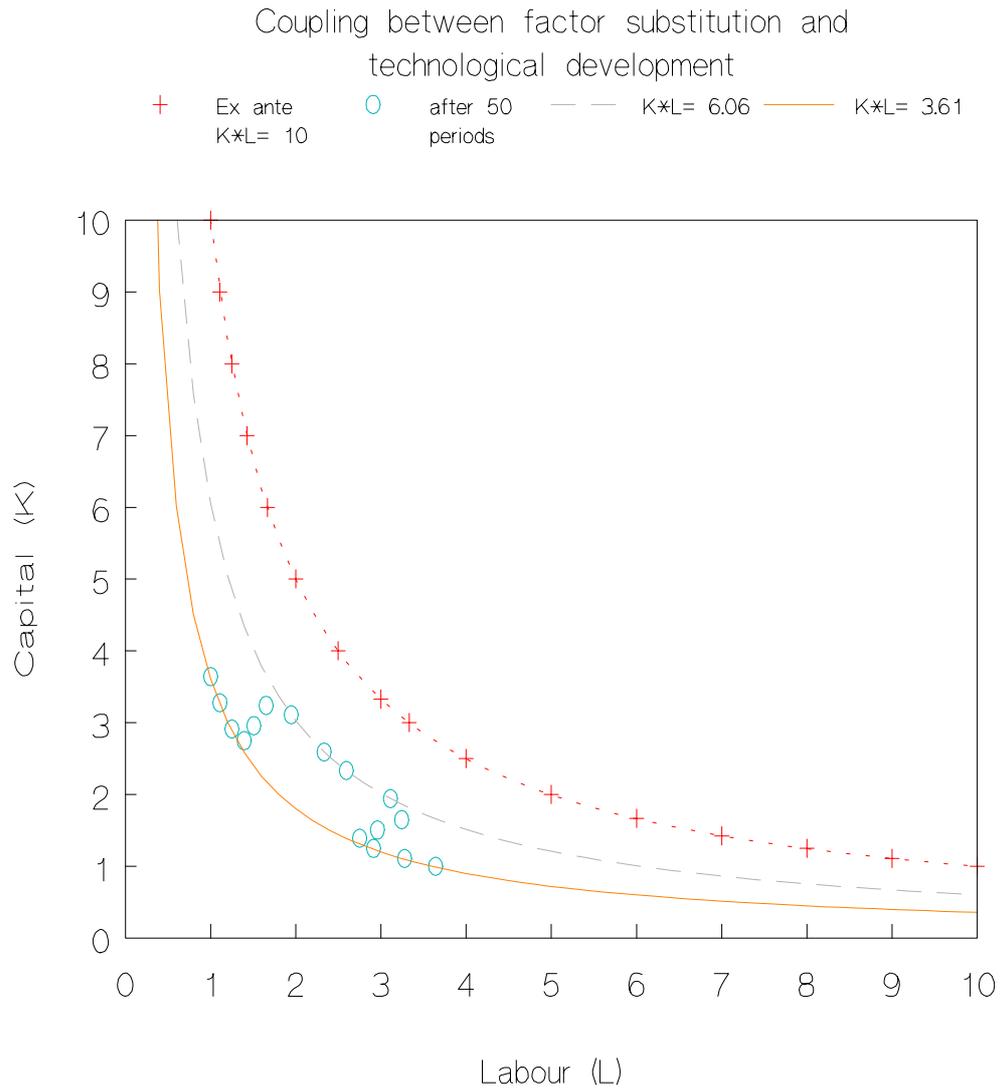

**Figure 3**

*Coupling between factor substitution and technological development leads to rigidities in the economy*

*Figure Three* exhibits the situation after 50 iterations. If this configuration is compared with *Figure Two*, we can see the emergence of two clusters of points. In this situation, however, it may be costly for an enterprise to move along the production function in accordance with changing factor prices, since this curve is no longer smooth. If, for one historical reason or another, enterprises choose an economic environment represented by one of these densities, they will be "locked into" a set of technologies to the relative exclusion of alternatives. The rigidities do not easily disappear, and may therefore influence further processes even if the historical reasons for the coupling have in the meantime disappeared.

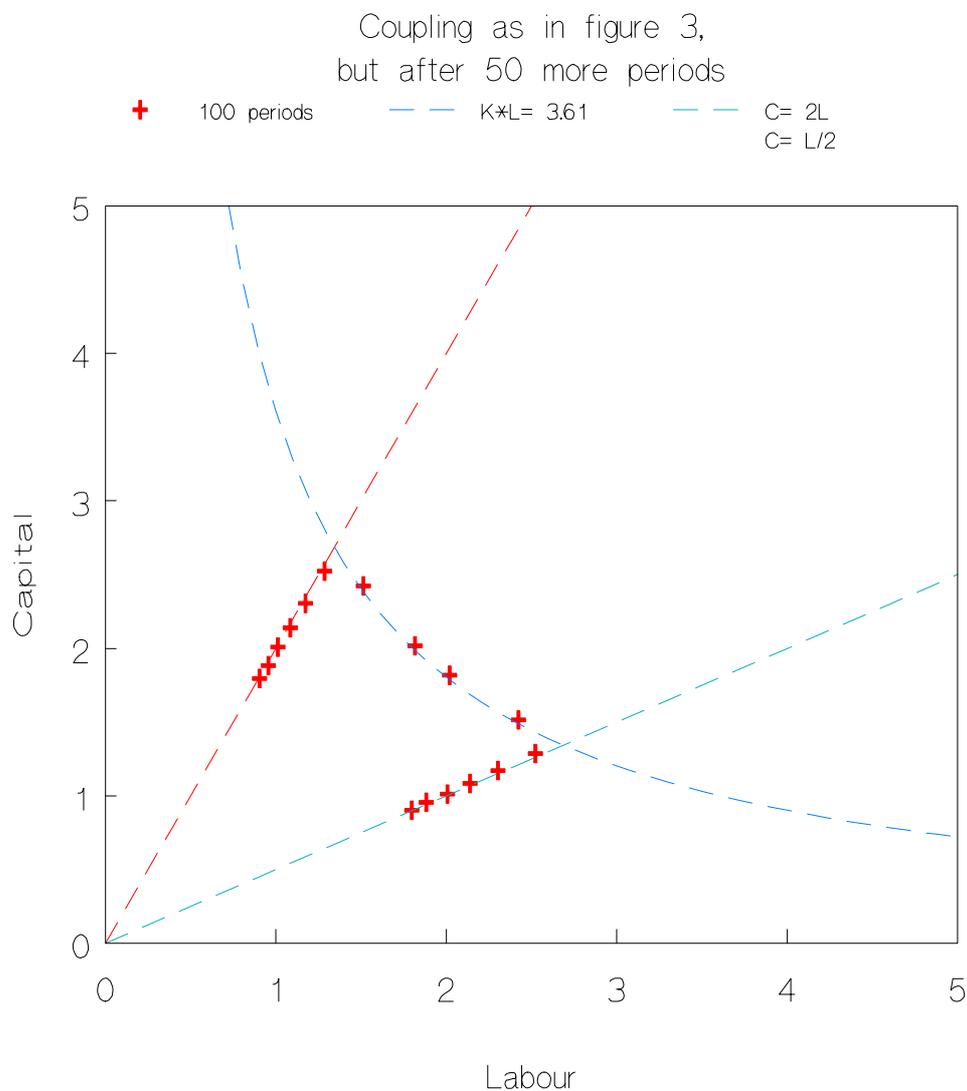

**Figure 4**

*Trajectory formation: coupling as in Figure 3, but after 50 more periods (n = 100)*

In summary, if different interactions are possible among factor substitution and technological developments under specific historical conditions, then this can lead to imbalances in the economic equilibrium. These imbalances can make it less attractive to choose other technologies despite changing factor costs, and thus the system may become locked into a specific state. *Figure Four* extrapolates the development shown in *Figure Three* for another fifty periods, i.e. under the assumption that the specific couplings continue to exist. In this case, the enterprises can eventually compete only in terms of the extent to which they have advanced along the technological trajectory for which they have opted.

**Negative feedbacks on possible savings of factor inputs**

In the above case, we assumed that factor substitution by technological development can be enhanced by relative factor costs. As a result, the originally linear reduction in factor inputs with one percent per period ($x_{t+1} = x_t - 0.01\ x_t$) was made more complex by assuming that this transition was no longer uniform in K (capital) and L (labour). For analytical reasons, we shall now first distinguish this interaction with factor inputs from a second-order feedback under the assumption of factor neutrality.

For example, if one assumes that production factors represent social forces, these forces can be expected to counteract their replacement by new technologies. Such "resistance" (e.g., by organized labour) can be represented as a one-higher-order feedback term in the previous model. For example, an additional feedback of one promille then leads to the equation: $x_{t+1} = x_t - 0.01\ x_t + 0.001\ x_t^2$.

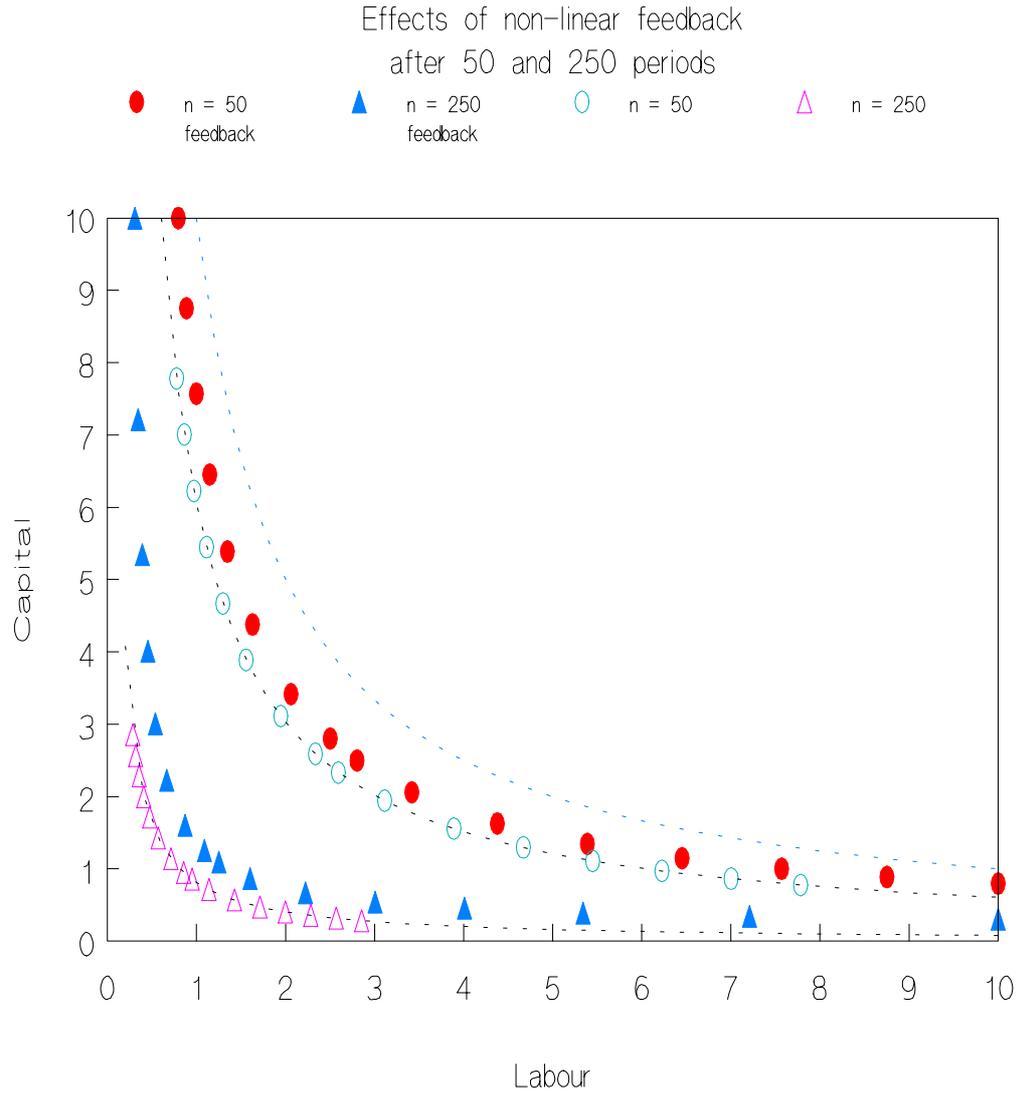

**Figure 5**

The results of this simulation are exhibited in *Figure Five*. They should be compared with those in *Figure Two*, representing the previous factor-neutral model. We see, first, that the movement of the production function towards the origin is indeed somewhat inhibited. But more importantly, we see that the range of options along the production function remains considerably larger. This effect can be socially important if, for example, one prefers to stimulate employment and technological developments at the same time.

In formal terms, one could say that the variance remains larger in the case of this feedback, since the non-linear term counteracts the first-order reduction of uncertainty. *Figure Six*, however, shows the risk in this configuration: we have increased the feedback term here to one percent, so that the coefficient for the linear factor reduction is equal to the one for the non-linear term (i.e.: $x_{t+1} = x_t - 0.01 x_t + 0.01 x_t^2$). In social reality, this corresponds with a much stronger resistance against reduction of factor input or with a decreasing pace of technological developments. In the model, the production function can move away from the origin in both these cases, i.e. the production processes become more expensive despite technological developments. The points spread along the axes with increasing speed. (We have sketched the situations after 15, 16, and 17 periods into *Figure Six* for the sake of illustration.) In other words: these production processes rapidly become more capital- or labour-intensive. The corresponding technologies exhibit not a stable, but an exploding trajectory; the situation tends towards crisis.[3]

---

3. The well-known bifurcation in the logistic curve ($x_{t+1} = ax_t - ax_t^2$) for values of *a* slightly larger than three, has no easy interpretation in this model, since the factor input has then to be more than triplicated in the linear term. In the economy, however, one assumes reduction of factor costs to be a dominant motive.

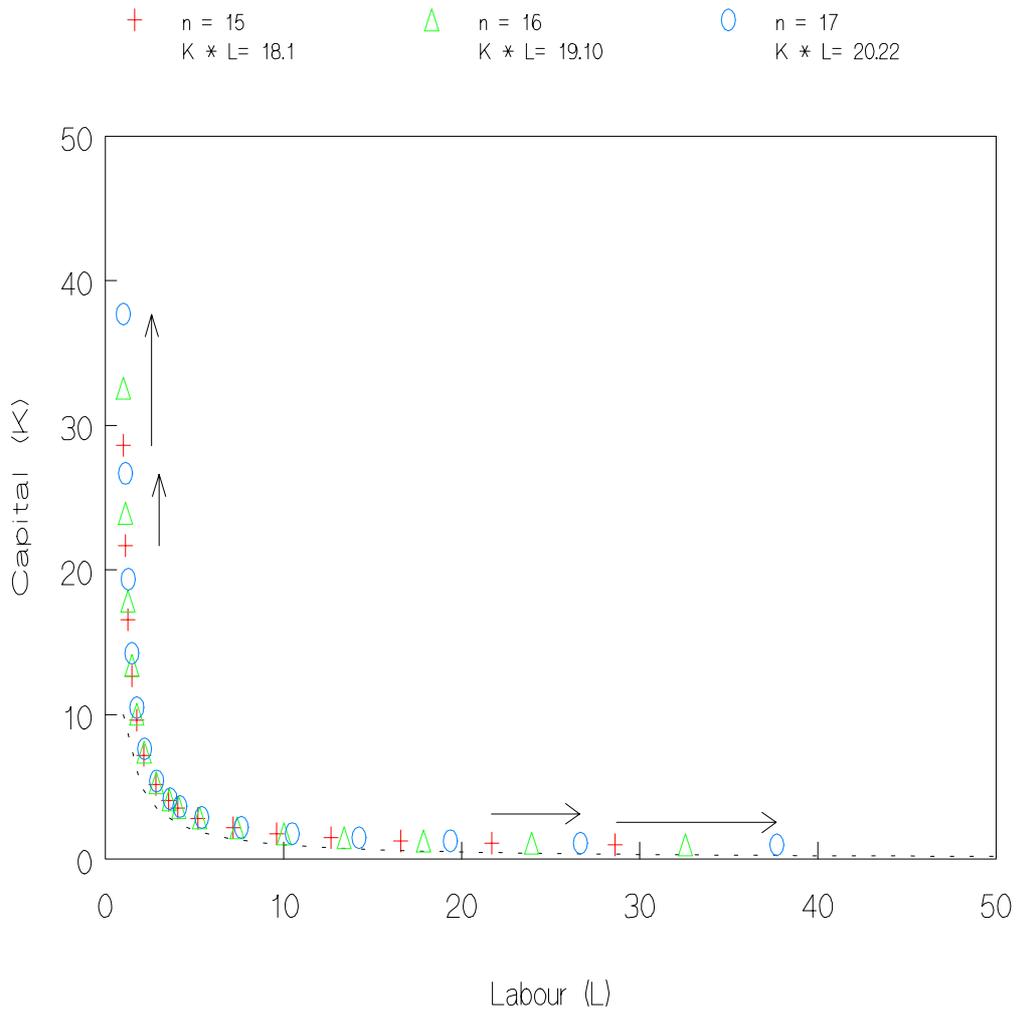

In the following sections, we shall argue that the social and institutional dynamics, represented above as a second-order feedback can be turned into a third-order feedforward in a more complex model; for example, when the factor expulsion by technological innovation can be reinforced as a result of scale effects. A technological regime is able to incorporate the institutional dynamics as one of its constitutive cybernetics.

**Technological trajectories, guideposts, and the possible emergence of regimes**

A comparison of *Figures Three* and *Four* suggests that one has to distinguish between states of the system before the attractor is reached (*Figure Three*), and states of the system after the "lock-in" (*Figure Four*). In the first stage, the system is under construction, and the trajectory exhibits a tendency towards an attractor.

Sahal (1985) has identified (hidden) invariants that are built into a system. He pictured his model as a mountain landscape with increasingly wide valleys, in which the sources located in the mountains represent "technological guideposts," while the valleys represent "innovation avenues." We suggest that in this stage the movement is *towards* stabilization; correspondingly, choices become more narrowly constrained with further evolution. Accordingly, the emerging "technological guideposts" can only be retrieved from an hindsight perspective.

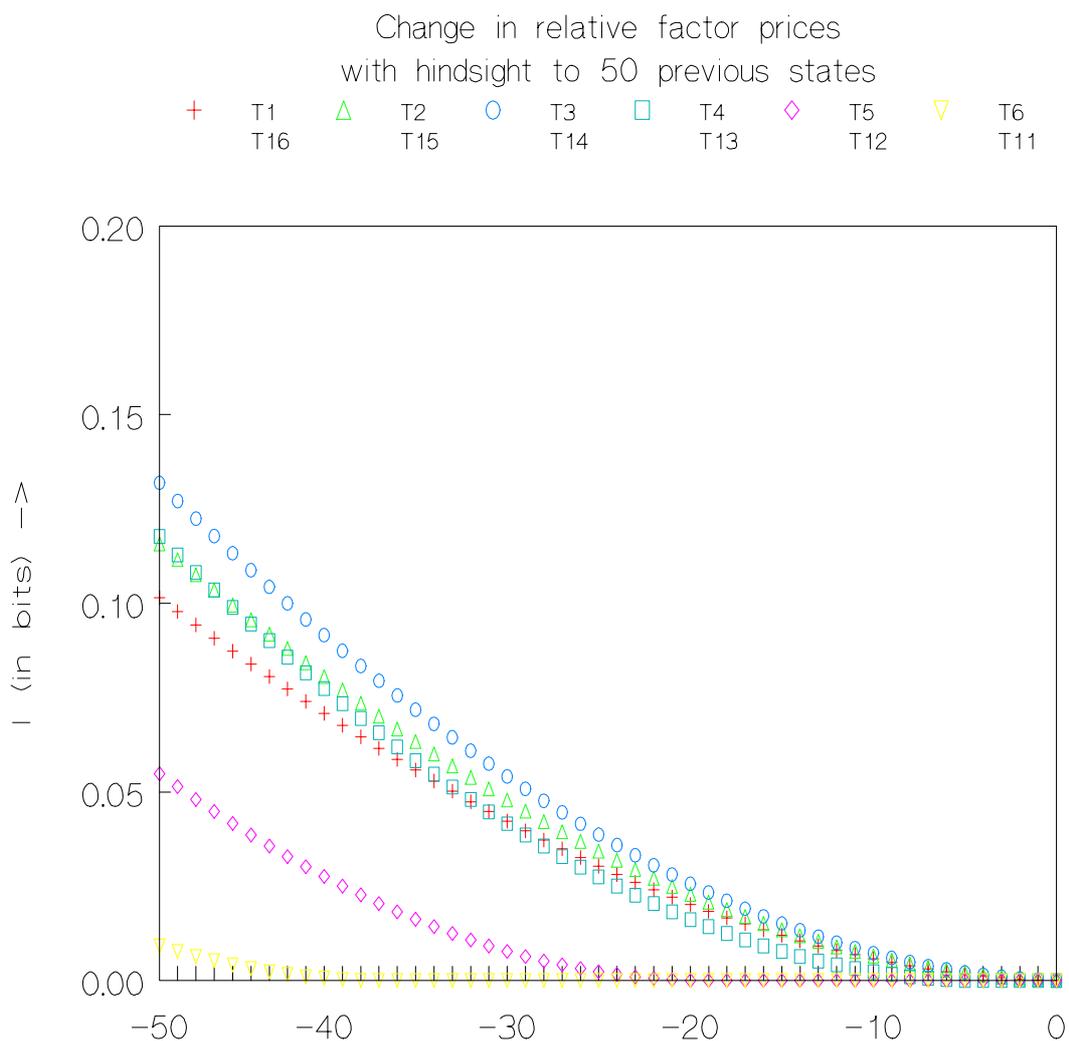

*Figure Seven* shows the consequence of this reversal of the time perspective in the evolutionary model: the various points in *Figure Three* are analyzed dynamically in relation to the fifty previous states of these technologies in terms of relative factor costs, by using Theil's (1972)

elaboration of Shannon's (1948) dynamic entropy measures for the comparison.[4] The figure shows, first, that there is a trend towards stabilization in all relevant cases, and second, that these developments differ depending on the starting configuration. The noted "invariants" are not determined by an *ex ante* configuration, but by the emerging *ex post* one. The trajectories and the regimes are evolutionary achievements.

As noted, the technological production processes may eventually be locked into an attractor (*Figure Four*). The technologies are then stabilized with respect to their capital/labour ratios; a stabilized technology transforms input into output according to a built-in production rule, and therefore can be considered as a given "black box" in the economic analysis. A stabilization, however, is by definition localized when represented on the production function. This localization can subsequently have an effect. As noted above, in the situation depicted in *Figure Three*, firms may increasingly have to choose between the two clusters of technologies, and this in itself may lead to concentration, scale effects, and "learning by doing" (cf. Rosenberg 1982).

These effects are localized, and consequently do not affect the various input factors equally. For example, scale effects are to be expected more on the capital side than on the side of labour. Let us, following this example, assume an additional percent of saving on capital during each period for the six points most to the left in *Figure Three*. We then obtain the points exhibited in *Figure Eight* after fifty and one hundred periods, respectively. In our opinion, this configuration can appropriately be termed *a technological regime*: alternative technologies are increasingly relegated to a marginal (i.e. economically unattractive) position. Thus, once technological trajectories are stabilized, there is no reason to assume a symmetric shift in the production function any longer since the selection environments have become specific. As a consequence the emergence of a superior single (set of) technologies is expected.

---

4. $I = \Sigma_i \, q_i * \log(q_i/p_i)$. *I* is the expected information content of the message that the *a priori* probability distribution $\Sigma p_i$ was changed into the *a posteriori* probability distribution $\Sigma q_i$.

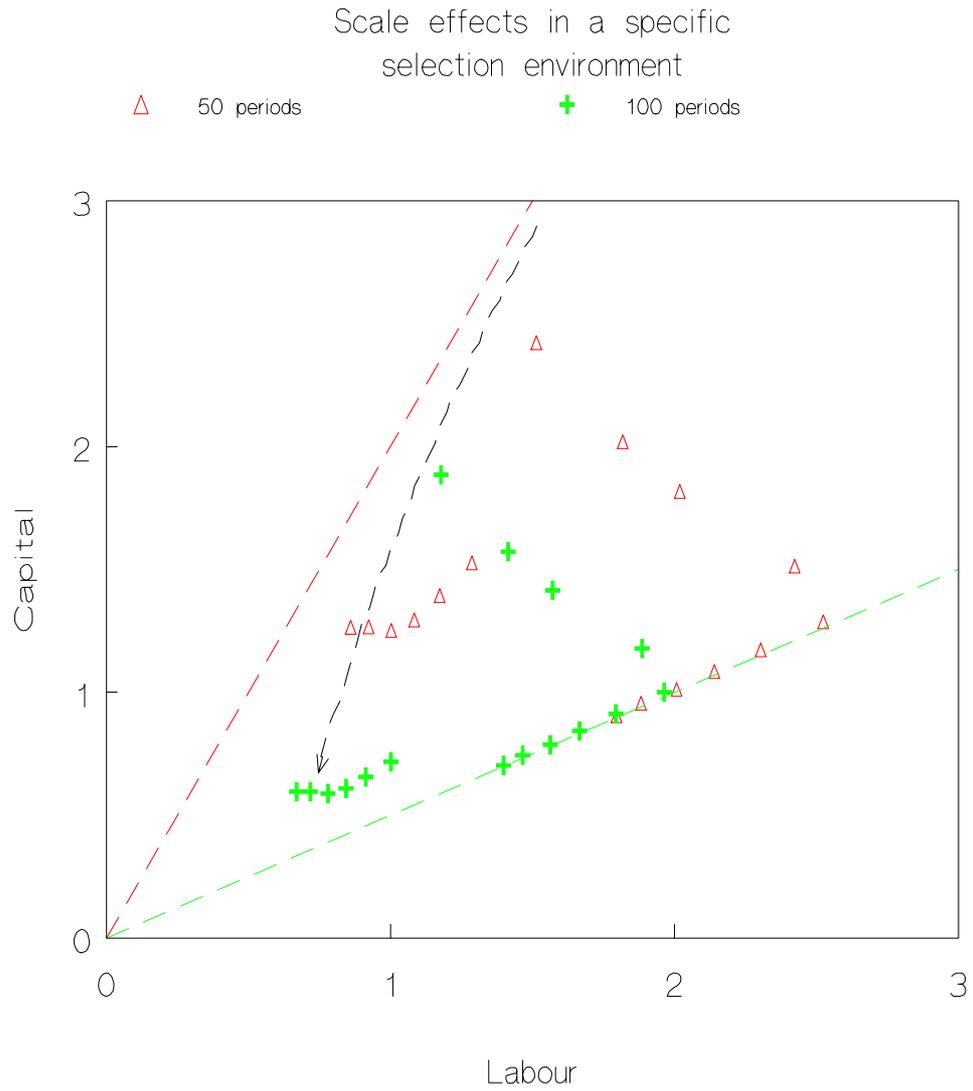

**Figure 8**

In other words, the different "sets of technologies" which can be represented in terms of various trajectories, have to compete in a higher-order cybernetics. When there is a specific environmental effect upon a set of technologies in one selection environment or another, this effect can lead to a technological regime. Note that a specific environment can be socially and institutionally supported (e.g., by a corporation or a network of institutions).

**The regime and its sub-cybernetics**

Our general argument has been that the higher-order dynamics finds its origin in the recursivity of the selective operation. When the variation in one context reaches a level of pronunciation (e.g., in terms of skewness), the co-variation can become so significant in another context that it can be considered a signal by this selection environment. The signal is transmitted in the network in terms of the substance that this network communicates when it operates. If the signal is subsequently received at some places, but not (or differently) at others, feedbacks are generated in the network system. The consequent possibility of a resonance induces the emergence of an attracting state (Simon 1969). The more complex system is *a posteriori*, and therefore this development is path-dependent and irreversible.

Of course, the complex system can again decay, but this is determined at the level of control of the more complex system, and only conditioned by the composing cybernetics. If the system has three degrees of freedom, it can be (provisionally) stabilized. In other words, stabilization means that the systems's reflexive function is fixed: input is converted into output according to a built-in production rule. If the system can develop in four dimensions, it is able to change additionally with respect to this rule, and therefore the system is able to learn, and exhibits a life-cycle of its own. The higher-order system, however, is more stable than the lower-level ones: if the higher-order system is in disarray, the lower-level system is less controlled, and therefore probably more active (as a muscle becomes hypersensitive upon denervation). Thus, this model can guide the search for clusters of innovations in the frequency domain (cf. Rosenberg and Frischtak 1984).

In the second part of this study, we wish to return to empirical questions by raising, for example, the question of the implications of the above reasoning for firm behaviour. How can one relate these results from simulations to observable phenomena? As noted, (distributed) firm behaviour is only one of the relevant sub-dynamics of the complex dynamic system that we have specified in terms of three sub-dynamics. Following Schumpeter (1939), we have *analytically* distinguished above between the equilibrating forces in the economy and the innovating mechanisms "upsetting the movement towards equilibrium." As a third mechanism, we have distinguished the organization of the social system. Under the assumption of a system of free enterprise, the latter can be analyzed in terms of a theory of the firm. At the level of society, one may think of technology assessment and other such studies feeding into the reflection on the social dimension (e.g., Schwarz and Thompson 1990).

Thus, three bodies of theory have been specified along three axes of the complex dynamic system, i.e.: (i) mainstream economics as the theoretical reflection on the price mechanism, (ii) evolutionary economics and innovation studies which add theoretical reflections on the mechanisms upsetting economic equilibria, and (iii) technology assessment and the theory of the firm which add reflections on the social and institutional dynamics of this complex system.

The fourth perspective is formal: the specification of the various sub-cybernetics generates a complex model that can be simulated algorithmically. The simulation enables us to evaluate the phase space spanned by the theoretical specifications, e.g., for alternative solutions. In general, the specification of theoretical insights on the basis of discursive interpretations is useful for a more precise specification of sub-cybernetics, and therefore theorizing may enable us locally to improve the model, i.e., in terms of one or more equations. The formal theory then enables us to proceed from theoretical specification toward generalization, in principle. Nowadays, this perspective is still rather programmatic (e.g., Anderson *et al.* 1988; Andersen 1994; Leydesdorff and Van den Besselaar 1994).

One expects additional theories to emerge concerning each of the interacting dynamics (cf. Simon 1973). This proliferation on the semantic side can lead to confusion if the relations between the substantive dynamics and the formal model are not sufficiently specified, and/or when the evolutionary models are used mainly in a metaphorical sense. For example, the relation between variation in the first dimension and selection in the second should be distinguished from the relation between "change" in the first dimension and stabilization over time as a third dimension. Selection (e.g., market clearing) is expected at each moment in time, but the sociologist may wish to use the metaphor of an engineer or an entrepreneur for the historical construction of a network (e.g., Callon *et al.* 1986; Bijker *et al.* 1987). Although the sub-dynamics are formally equivalent, they have profoundly different meanings.

**Specification of the dynamic interactions between trajectories and environments**

As noted, Nelson and Winter (1977, pp. 61 ff.) defined selection environments as *dynamic boundary conditions* to (in themselves probabilistic) trajectories. Since these authors (*ibid.*, p. 49) did not allow for (higher-order) feedbacks between environments and trajectories, the dynamic boundary conditions for the technological developments could be specified *ex ante*. When technological trajectories stabilize, they increasingly feedback on their own boundary conditions, and thereby the transition probabilities will be changed. Therefore, one can understand the *ex ante* specification of a constraint as one possible type of mutual conditioning between technological trajectories and selection environments. The conditioning *ex ante* is exclusive only while technological trajectories are in the construction phase. However, other interactions are possible when the nature of the interaction has changed over time.

For analytical reasons subsequent dynamic redefinitions of the relations between two independent variables have to be generated on the basis of developments in either system or on the basis of their interaction. In formula format:

$$F(x,y)_{t+1} \quad = \quad ax_t^{\alpha} + by_t^{\beta} + c(x_t^{\gamma} \cdot y_t^{\delta})$$

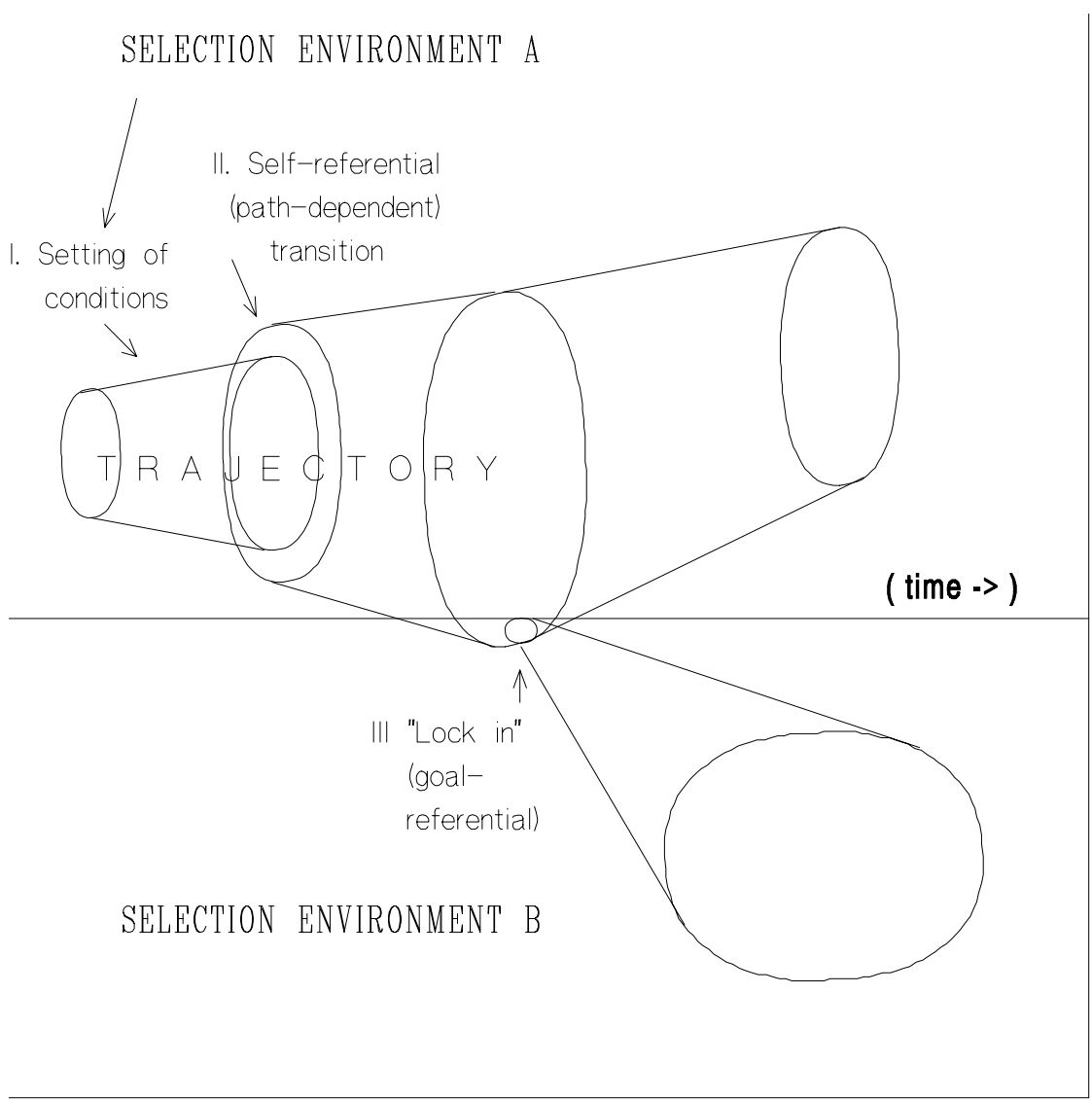

Three options for dynamic definitions and redefinitions of boundary conditions between subsystems follow. *Figure 9* may serve as an illustration (Leydesdorff 1992). First, the one (sub-)system may *ex ante* condition the developments in the other one, as discussed by Nelson & Winter (1982). Second, when either (sub-)system develops internally, and reaches a new state by building on its previous state, this internal evolution can drive a redefinition of the relations with this (sub-)system's environment (Sahal 1981). Third, the one (sub-)system (represented by x) may drift into a different relation to its environment (y) without necessarily having gone through an internal path-dependent transition (cf. Arthur 1988). In this case, change originates from passing a threshold which represents a selective operator that is inherent to the other system (cf. Von Hippel 1988).

These three options correspond to three existing models of the innovation process, if the latter is conceptualized as a result of possible interactions between economic demand and technological development. Demand can be specified either in reaction to technological options which have become available thanks to path-dependent technological developments ("technology push"), or it can be specified (analytically) independently, e.g., in the economy. In the latter case, the specified demand may have driven the relevant technological activities, and thus have been part of the *a priori* conditions of the probability distribution of the transitions in the technological system ("demand pull"), or it may have been "met" by technological developments which were not yet conditionalized as a function of this demand. In the latter case, the system has drifted into a "lock-in."

The three mechanisms are based on different assumptions. While internal change requires specification of a transition matrix for one system, a "lock-in" requires specification of the relevant environmental thresholds in the other system. In the case of technological developments, these latter thresholds are determined by dynamic demand functions: when does a technological system generate a signal that resonates in a market? However, these demand functions have then to be specified, since otherwise one would have no criterion for assessing whether or not "demand" was met (as a threshold). Note the change in analytical perspective from the *ex ante* specification of boundary conditions on technological developments in the model proposed by Nelson and Winter (1982): in the case of a lock-in the analyst specifies what the market (or the selection environment) is expected to consider as a significant signal from the side of the technological system, and not what the latter is to consider as its condition by the former.

As a fourth mechanism the new state can also be a consequence of longer-term adjustments between the two systems x and y, instead of the passage of a threshold at a specific moment in time. When this occurs, one obtains a co-evolution model between specific technologies and selection environments (cf. Nelson 1994). Co-evolution, however, requires gradual stabilization of the interaction term (xy) between the systems x and y into a super-system that can feed back on both variation (e.g., x) and selection (e.g., y). Upon stabilization of the co-variation between x and y, the interaction term can be considered as a third independent variable ($z = f_{(xy)}$), of which the path-dependent transitions in relation to x and/or y can be described at next-higher level, but in similar terms.

The operations are recursive, but the higher-order operation requires one more degree of freedom than the lower-level one. Thus, the emerging variable (z) can only co-vary as an independent source of variation with either of the other (sub-)systems x and y, if the third axis has previously reached a position orthogonal to the ones that contributed to its construction. This third degree of freedom, however, corresponds to a third system of reference that should then be provided also with a theoretical appreciation.

In the orginal (neo-classical) representation the two dynamics of factor substitution and technological development were represented as orthogonal dimensions. It has been our argument that developments in otherwise orthogonal dimensions can co-vary, and that the system can gain in complexity if this co-variation can be stabilized and internalized as an additional dimension of the system. For example, the interactions between technological trajectories and economic environments can be conditioned by organizational factors (Van den Belt & Rip 1987). These institutional dynamics can be carried by an entrepreneur (e.g., Hughes 1987), a sector (e.g., Abernathy and Clark 1985; Nelson 1982; Pavitt 1984), or an interorganizational network (e.g., Clark 1985; Shrum 1985). We have shown above that if this third context is additionally incorporated, this can lead to the co-evolution of a trajectory and an environment into a technological regime.

Trajectories can be stabilized in three dimensions, and can therefore be observed in terms of firm behaviour. The technological regime, however, integrates over time three (nearly decomposable) dynamics into a hyper-dynamics that can no longer be fully understood by using geometrical metaphors. By understanding variables as fluxes, the algebra allows us to distinguish between change in the value of variables and change in the categories themselves. Discursive theories specify sub-cybernetics by stabilizing a specific reflection on the observed system in the scholarly communication (Hinton *et al.* 1986; Langton 1992, pp. 22 ff.). Each discourse offers a window of appreciation on the more complex dynamic systems under study.

**Implications for firm behaviour**

In order to specify the implications for firm behaviour, one has to disaggregate the distributed context of firm behaviour, and to examine the possible reflections of the interaction between the other two sub-dynamics (trajectories and environments) on this third dimension. Firms differ among them in terms of their position. Therefore, the choice of each focus raises empirical questions about the range of possible reflections. These empirical questions go beyond the scope of the present study (e.g., Pavitt 1984; Faulkner and Senker 1994). However, we are able to indicate which patterns of firm behaviour one expects on the basis of our argument.

In general, the dynamics between two dimensions of a complex system can be expressed in formal terms using the equation given in the previous section: in each case, two independent variables (x and y) are assumed that can at a a next moment (i.e., at *t+1*) both have interacted ($x_t^\gamma \cdot y_t^\delta$), and/or have built on their previous states ($x^\alpha$ or $y^\beta$), but the latter *only if* these systems had previously been stabilized. Both independent variables may thus represent the stabilization of previous developmental processes ("continuity") and/or contingent "change."[5] Three types of interactions are then conceivable: first, both independent variables have been contingent; second, one independent variable is contingent while the other represents a stabilization of prior developmental processes; or third, both independent variables represent a stabilization of prior developments. Let us specify the expected developments:

ad 1. When both independent variables are contingent, the processes characterizing the development are those of variation and stabilization, since there has been no structural basis for selection. In view of its path-dependent nature the (stochastic) variation will eventually be locked into a stabilization (Arthur 1988). At the organizational level, one expects in this case a flexible organization which takes its chances for system building, and hence pushes for stabilization. As noted, the dominant metaphor is that of an engineer or entrepreneur who constructs the system.

---

5. The relation between auto-correlation and change provides us with options for testing the stability of the system (cf. Leydesdorff 1991).

ad 2.   When both contingent and previously stabilized variables are involved in the interaction, the processes to be expected are those of variation (introduced by the contingent variable), selection (introduced by the stabilization of prior development processes into structures), and stabilization.  Furthermore, since the previously stabilized factor--by definition--refers to its previous state, feedback loops characterize the dynamics and introduce, depending on the sign, either self-amplification or self-dampening (towards equilibrium).  More specifically, the alteration in the sign of the feedback is expected to lead the life-cycle of the technology (cf. Abernathy and Clark 1985).  The organizational emphasis should be on innovation in the upswings, and on cash-flow in the down-swings.  Therefore, one expects a more complex configuration with diversified profit centres in order to profit fully from these alterations.  The dominant focus of scholarly attention is in this case on the industry.

ad 3.   When both independent variables represent prior stabilizations, the relevant processes are, theoretically, those of selection and stabilization.  In this case, the introduction of contingency requires an organization as a third dimension in which one is able to generate "newness" from interactions among selections made on either side.  As noted, Nelson & Winter's (1982) "search and selection processes" contain such dual selections.  The question of stabilization then leads to the question of whether, and for how long, this interface can be sustained.  The repetition of the co-variation can lead to co-evolution if the specific interactions can be stabilized.

Note that these analytically distinguished processes do not exclude one another, but one expects the various cycles to be of a different order.  The empirical assessment of the relevant contexts in terms of the stability of their operation is therefore of prime importance in model building which seeks to provide normative advice (Dosi 1991; Brunner 1994).

**Summary and conclusions**

Since technological innovation is not inherently factor-neutral, economic rigidities are developed: technological trajectories emerge *in the economy*.  If certain trajectories offer specific advantages to entrepreneurs, one may expect the production system to be "localized" increasingly in this area of the production function.  Densities in the network, however, are initially latent for the actors involved (Lazarsfeld and Henry 1968). The actors are positioned in terms of their history, and can only gradually develop in the direction of perceived optima (cf. Sahal 1981).

Nelson & Winter's (1982, p. 19) observation that "(t)hrough the joint action of search and selection, the firms evolve over time, with the condition of the industry in each period bearing the seeds of its condition in the following period" highlights the crucial role of a possible co-evolution between the two network functions of cost reduction and innovation in a specified organizational context. If this third dimension is additionally stabilized into the complex system, a self-organizing regime may emerge (since time provides us with the fourth dimension).

If a regime is established as a more complex system, it is expected to exhibit a cycle of its own, and consequently to transform the economy. The system may lock into attracting states, but the regime remains subject to the price mechanism, technological learning, and organizational learning (e.g., scale effects) as its constitutive cybernetics. For evolutionary reasons, the higher-order system emerges nearly decomposable in functional sub-systems (Simon 1969). Therefore, one may expect the generation of reflexive theories that specify the mechanisms of the various sub-systems, and additionally, theories which focus on the interactions between the various composing dynamics (Simon 1973).

We have distinguished above among (i) mainstream economics as the theoretical reflection on the price mechanism, (ii) evolutionary economics and innovation studies as theoretical reflections mainly on the mechanism upsetting economic equilibria, and (iii) technology assessment and the theory of the firm as the primary reflections on the social and institutional organization of this complex. These distinctions have enabled us to specify further reflections which focus on the various interaction terms.

The problem of specifying the likelihood of developments remains discursively tractable for the case of relations between two dynamics. Various instances of two-factor designs were specified, and we have shown their correspondence with the various theories about innovation. As soon as more than two dynamics are involved, the specification of the probability distributions for changes (including innovations) requires an algorithmic approach. Consequently, in the case of a technological regime, innovations occur where one does not expect them intuitively or on the basis of discursive reasoning.

**Policy implications**

We have focused above on normative implications for firm behaviour. Analogously, implications can be elaborated for government interventions at the system's level (cf. Nelson 1993). In general, the (often unintended) function of technology policies in advanced societies is to disturb the development of existing regimes. Such disturbances, however, are expected to reinforce these regimes. While the regime (e.g., the car system) can be expected to restore its own order, taxing leaded petrol may be helpful for developing catalysts or cleaner engines as new trajectories. Technology policies can be particularly effective at the level of the sub-systems (e.g. trajectories) if they provide the relevant agents with a new context for the "creative destruction" of a specified hyper-system. As noted, intervention at the level of each sub-cybernetics can be stimulative at the level of the larger system, but only in an *ex ante* uncontrollable way.

From this perspective, the debate over socialist intervention and liberal *laissez-faire* has grown obsolete: today's regimes contain both the dynamics of markets and those of social organization. Nowadays, the crucial question for technology policy is whether interventions will prove to be functional in disturbing the ongoing processes by generating innovations without laming other sub-cybernetics. It was shown (for example, in *Figure Six*) that these delicate balances can be expressed only probabilistically.